\documentclass[letterpaper, 10 pt, conference]{ieeeconf}  

\IEEEoverridecommandlockouts                              
\usepackage{graphics} 
\usepackage{epsfig} 
\usepackage{amsmath} 
\usepackage{amssymb}  

\usepackage{url}
\usepackage[ruled, vlined, linesnumbered]{algorithm2e}
\usepackage{verbatim} 
\usepackage{soul, color}
\usepackage{lmodern}
\usepackage{fancyhdr}
\usepackage[utf8]{inputenc}
\usepackage{fourier} 
\usepackage{array}
\usepackage{makecell}
\usepackage{todonotes}
\usepackage{hyperref}

\usepackage[
backend=biber,
style=numeric-comp,
sorting=none
]{biblatex}

\SetNlSty{large}{}{:}

\makeatletter

\newcommand{\Rom}[1]{\expandafter\@slowromancap\romannumeral #1@}
\makeatother

\title{\LARGE{ \bf{
Recasting the ATLAS search for displaced hadronic jets in the ATLAS calorimeter with additional jets or leptons using surrogate models}}}

\author{\parbox{5 in}{\centering Louie Dartmoor Corpe$^1$, Abdelhamid Haddad$^1$, Mark Goodsell$^2$\\
\vspace{1cm}
${^1}$\textit{Université Clermont Auvergne, CNRS/IN2P3,  Laboratoire de Physique de Clermont Auvergne, 63000 Clermont-Ferrand, France}
\\
${^2}$\textit{Sorbonne Université, CNRS, Laboratoire de Physique Théorique et Hautes Énergies (LPTHE), 75252 Paris, France}}
}

\begin{document}

\maketitle
\pagestyle{plain}

\begin{abstract}

This note describes the validation of a new form of re-interpretation material provided by an ATLAS search for hadronically-decaying neutral long-lived particles in association with jets or leptons, using the full Run-2 dataset. This reference ATLAS analysis provided a set of machine-learning-based ``surrogate models'' which return the probability of an event being selected in a given channel of the analysis, using as input truth-level kinematic information (decay position, transverse momentum and decay products of the long-lived particles). In this document, we describe the surrogate model framework in detail, and how it responds to issues identified in other re-interpretation procedures. We describe independent validations of the surrogate models’ performance in reproducing the original analysis results — first using a standalone framework and then employing the HackAnalysis framework.\\

\end{abstract}

\section{\textbf{Introduction}}
\label{sec:intro}

Long-lived particles (LLPs) are predicted in many extensions of the Standard Model (SM) of particle physics, and could have been missed by the traditional search programme~\cite{LLP_white_paper}, which is not optimised for displaced activity. Hence, dedicated searches are often needed, when searching for LLP decays. A great amount of progress on these and other unusual signatures has taken place over the last half-decade, often relying on dedicated reconstruction, exploiting the response of the detector to unusual signatures, and making heavy use of machine-learning techniques.

However, there is a downside to this novel exploitation of the detector. LLP searches often suffer from very low re-interpretability, since people outside the collaboration do not have access to the detailed response of the detector, and hence are typically unable to replicate the selections of the original analysis. So, at best, the benchmark models used in the original analysis can be embedded in a more realistic model. However, if one moves to a model with different kinematics, then it's almost impossible to predict accurately how the selection efficiency would change.
 The ``single-use'' nature of LLP searches is a drawback to their long-term impact on the field. This ``LLP re-interpretation challenge'' must be resolved to ensure current LLP analyses have impact far into the future.  This issue has motivated the development of new techniques in analysis preservation, which in turn can have wider applicability in other machine-learning based searches. A series of searches for displaced hadronic jets, in particular, have led the charge in developing new material. Indeed, over the years, they have evolved to give ever more complete recasting tools:
 \begin{itemize}
     \item Ref.~\cite{ATLAS:2015mlf} provided just numerical data for limits (allowing a new model to be constrained as long as the LLP sector remains intact with respect to the original analysis benchmark model); 
     \item Ref.~\cite{ATLAS:2019qrr} offered a set of per-event efficiency curves as a function of LLP lifetime for different new particle mass assumptions (which allows re-interpretation in terms of new models as long as the LLP kinematics do not change too much);
     \item Ref.~\cite{EXOT-2019-23} gave efficiency-map objects that link the truth-level LLP kinematics and decay positions to a probability of the event being selected (which, in principle and within some conditions, allows one to reproduce the selection efficiency for the original benchmark model and provides an approximate efficiency for new models with different kinematics).  The performance and limitations of the this method after an independent validation were described in Ref.~\cite{ThomasNote}.
\end{itemize}
Most recently, in the latest iteration of displaced hadronic jet searches~\cite{ATLAS-EXOT-2022-04}, the aforementioned efficiency-map objects have been promoted to machine-learning models which we refer to as ``surrogate models'' (SuMos),  intended to be more performant, user-friendly and portable than the previously-provided recasting tools. This means, for the first time, the recasting materials fulfil the FAIR (Findable, Accessible, Inter-operable and Re-useable) principle of open science. In this note, we present an independent validation of the SuMos provided in Ref.~\cite{ATLAS-EXOT-2022-04}, evaluate their performance when trying to reproduce the results of the original analysis, and identify remaining areas for improvement.

 \section{\textbf{Description of the reference analysis}}
\label{sec:refana}

Before describing the SuMos, we first provide a short summary of the analysis selections which they are trying to approximate. The analysis in question is the ATLAS search for hadronically-decaying LLPs in the calorimeters, produced in association with a prompt (leptonically-decaying) $Z$ or $W$ boson, or in association with additional jets~\cite{ATLAS-EXOT-2022-04}. It follows a similar structure to a previous search for pair-produced LLP decays in the calorimeter~\cite{EXOT-2019-23}, which was optimized for cases where the decay products from each LLP (for instance two $b$-quarks) were very collimated, leading to a single ''merged'' jet in each case. This analysis is sensitive for cases where the LLPs have high transverse energy, but breaks down in regimes where a large fraction of LLP decays can lead to two resolvable jets. Hence, Ref.~\cite{ATLAS-EXOT-2022-04} was a re-optimised version of the analysis to allow one of the jets to be ''resolved''. This channel is named ''CR+2J'' in reference to the fact that the displaced jet yields an unusual calorimeter energy ratio (CR) and is accompanied by two resolved jets (2J). The same paper additionally studies cases where the LLP or LLPs were produced in association with a $W$ or $Z$ boson, with several selections optimised for different kinematic regimes. Those channels are dubbed ''CR+$W$'' and ''CR+$Z$'' respectively.

As in previous similar searches, a template model based on the Hidden Abelian Higgs Model (HAHM)~\cite{Wells:2008xg,Curtin:2013fra} is used as one of the primary benchmarks. The HAHM introduces three new particle: scalars $\Phi$ and $S$, and a dark photon $Z_d$. In the configuration used in previous analyses, the couplings to the $Z_d$ are set to zero, and the LLP $S$ is produced via a scalar mediator $\Phi$ which can be the Higgs boson or some other new particle. This configuration is called the ''Hidden Sector'' (HS) reference model. The mediator can be produced via any of the usual Higgs boson production mechanisms, but the reference analysis is optimised for gluon--gluon fusion or associated vector boson production. The mediator then decays to a pair of long-lived scalars $S$  which eventually decay back into the SM via a Yukawa interaction. This model is probed in all three channels of the analysis. Another configuration of the HAHM decouples the $S$ particle, and instead has the mediator decaying to a long-lived dark photon $Z_d$ and a regular $Z$. This configuration is studied in the CR+$Z$ channel. Finally, The CR+$W$ and CR+$Z$ channels study a model where long-lived axion-like particles (ALPs) are radiated from a vector boson and decay exclusively to gluons~\cite{Brivio:2017ije}. Illustrative Feynman diagrams for each of these models can be found in Ref.~\cite{ATLAS-EXOT-2022-04}.

The full LHC Run 2 dataset is used to look for pairs of displaced jets originating from hadronic decays of neutral LLPs in the hadronic part of the calorimeter. 
The displaced jets are expected to be largely trackless (since the LLP is neutral) and with a high proportion of energy in the hadronic part of the calorimeter with respect to the upstream electromagnetic calorimeter (since the search focuses on decays in the hadronic part). These selections are helpful to reduce the QCD multijet backgrounds. The data were collected with dedicated LLP triggers which also exploit this calorimeter energy ratio topology, or standard lepton triggers in the CR+$W$ and CR+$Z$ channels. 

The search makes use of a neural network (NN), which was developed initially for Ref.~\cite{EXOT-2019-23}. The inputs include deposits in each layer of the calorimeter as well as tracker and muon system information. It classifies jets as likely to have come from QCD processes, signal, or beam-induced background (such as beam halo muons which travel in parallel to the LHC beam and strike the detector without passing through the interaction point). 

There are six selection in all: one for the CR+2J channel; two selections targeting the CR+$Z$ channel, optimised for high- and low-energy transverse energy regimes (ZHS\_highET and ZHS\_lowET); and three selections targeting the CR+$W$ channel in different regimes (WHS\_highET, WHS\_lowET and WALP, with the latter focused on extremely light particles). All these selections make use of event-level machine-learning tools exploiting the NN scores for the jets most likely to originate from a signal process, along with additional event-level information, to classify events as signal-like or not. The CR+$W$ and CR+$Z$ channels make use of prompt leptons and missing transverse momentum to reconstruct and select vector boson candidates. Some additional cleaning requirements (on signal-like jet timing, momentum and position, etc...) are used to remove events likely to have originated from cosmic ray muons or beam-induced background. 

The event-level machine-learning score in each channel is then used with an orthogonal variable measuring the proximity of tracks to the jets in the events, to form ABCD planes with which data-driven background estimates can be made. Some differences exist between the strategies of the various channels, but the overall methods remain the same. The signal is mostly concentrated in region A of the plane, while the background (which at this stage is composed purely of QCD multijets or $W/Z$+jets processes) is distributed in the other regions in an uncorrelated way. A simultaneous fit of the signal yields and background yields (which should obey the $A=B\times C/D$ relation) is performed for the final statistical analysis. The signal efficiencies in region A are extrapolated to arbitrary lifetimes using an exponential reweighting method.

\section{\textbf{Re-interpretation material description}}

\subsection{\textbf{Structure}}

The reference analysis provided re-interpretation material in its HEPData record~\cite{HEPData} which includes digitized limit curves and event-level efficiencies as a function of LLP proper decay length times the speed of light  ($c\tau$). Under ''Resources'', it also points to a zenodo record~\cite{Zenodo} containing the aforementioned SuMos (in pickle and ONNX~\cite{onnxruntime} format), along with some template code and sample events to test them with.  

The SuMos are implemented as multi-class Boosted Decision Trees (BDTs). Specifically, they are \texttt{RandomForestClassifier} objects from the python SciKitLearn (v1.4.2) library~\cite{scikit-learn}. This widely-used format has the advantage of being highly portable, in particular because it can be evaluated by the industry-standard ONNX runtime package~\cite{onnxruntime} in a variety of programming languages including C++ and python. This facilitates integration of the SuMos in existing re-interpretation tools without the need for a custom implementation.
For example, the SuMos have already been integrated in the \textsc{HackAnalysis}~\cite{Goodsell:2024aig} framework, and some validation plots from that tool are included in this note in Section~\ref{hackana}.

\subsection{\textbf{Inputs and outputs}}

Each SuMo takes as input truth-level kinematic information about the LLPs and associated prompt objects. It returns a list of five numbers (between 0 and 1) which represent respectively the probability that the event would enter five mutually exclusive classes: not be selected, selected in region A, selected in region B, selected in region C, and selected in region D.  The SuMo objects themselves can be thought of as ``forward folding matrices'', which encapsulate the effects of the detector and the analysis at once. They rely on the principle that the detector (and more broadly, the analysis) cannot be sensitive the the internal details of the model, but only to observable properties of the outgoing particles, such as the decay position, kinematics and decay types (codified through the decay product particles' PDGID). Each selection in the analysis has a separate SuMo to approximate it, with a list of input features which reflect the topologies targeted in each case. For example, the CR+$Z$ and CR+$W$ selection SuMos take the following inputs: 
the transverse and longitudinal decay positions, transverse mass\footnote{The input feature list for the SuMos originally incorrectly stated this variable as transverse energy, but in part thanks to this validation, it was realised that transverse mass was the actual input feature.}, transverse momentum, pseudorapidity and decay type of each of the two LLPs, as well as the transverse momentum and pseudorapidity of the vector boson. The CR+2J SuMo takes the same inputs but without the vector boson kinematics, since none are expected in that channel. The dimensionful inputs should be provided in units of GeV or meters as relevant. The inputs to the BDT need to be pre-scaled, a routine practice in machine-learning. Specifically, for each input feature $X$ of each SuMo, two constants are provided: a mean $\Bar{X}$ and standard deviation $\sigma_X$. The scaled value $X'$ can then be obtained from $X' = (X-\Bar{X})/ \sigma_X$. The mean and standard deviation constants are provided in files ending with \texttt{"\_scaler\_mean.npy"}  and \texttt{"\_scaler\_std.npy"} respectively. \textbf{The ONNX files for each SuMo already perform this pre-scaling step}. Hence, this only needs to be done manually if using the SuMos stored in pickle format. Examples of how to do this prescaling correctly are provided in the HEPdata record for the analysis.

 A final consideration is that certain selections in the CR+$W$ and CR+$Z$ channels are sensitive both to models where a single LLP is present and to others were two LLPs are present. The SuMos however do not handle a variable number of LLPs and always expect the inputs for two LLPs. In the case where a single LLP is present in the model, the recommended procedure is to fill both sets of LLP inputs with the same values. In the case where two LLPs are present, one should randomly select which is LLP1 and which is LLP2 in the input list.

\subsection{\textbf{Usage}}
 
To obtain the selection efficiency for a given sample of generated events for a given selection, one should evaluate, for each event, the five BDT class scores using the \texttt{predict\_proba()} method if using the pickle file or selecting the \texttt{output\_probability} output from ONNX runtime. One should then sum the scores across all events in the sample and divide them by the total number of events (adequately weighted if applicable). This then provides five approximated efficiency values, one for each class. The most important class is typically the ''selected in Region A'', which is the 1st in the list (rather than the 0th, which is the ''not selected'' class).

\subsection{\textbf{Uncertainties and bounds of applicability}}

The helper code which accompanied the HEPData material suggests that the estimated efficiencies are typically accurate within 25\%. This claim is borne out by the validation performed in this note. Further, the helper code also suggests bounds of applicability where the approximated efficiency values can be trusted. Specifically, the results can be trusted when the average transverse decay position of the LLPs in the barrel ($|\eta| < 1.5)$ region is between 25~cm and 16~m, and the longitudinal decay position of the LLPs in the endcaps is between 75~cm and 28~m. These bounds represent cases where an appreciable fraction of the decays happen in the calorimeters. For samples where these constraints are not met, the results should be treated with caution as they may overstate the efficiency.


\section{\textbf{Validation using standalone framework}}

In order to validate the performance of the SuMos, we generate samples from the models used in the reference analysis with matching mass parameters, and seek to reproduce the efficiencies which are provided in the Auxiliary Material of the paper and on HEPData for a wide range of parameter values. Specifically, we use MadGraph~v3.4.2~\cite{Alwall:2014hca} interfaced with Pythia~8~\cite{Bierlich:2022pfr}  to generate HEPMC~\cite{Buckley:2019xhk} events for an arbitrary lifetime. All MadGraph and Pythia settings are chosen at their default values, and the \textsc{NNPDF2.3LO} PDF set is used. To evaluate the efficiency as a function of $c\tau$, we treat this parameter as independent from the other model parameters such as mass and coupling, and instead sample an exponential distribution for the proper decay length, which we convert to a lab decay position by applying a Lorentz boost dictated by each particle' kinematics. The UFO files, \textsc{MadGraph} cards and processing code are stored in Ref.~\cite{recastcode}. The event generation setup is very similar to the one used in Ref.~\cite{thomascode}. For each sample, 20~000 events are generated.
The efficiency for each sample and for each lifetime is obtained by summing the probabilities estimated by the SuMo for each event and dividing by the total number of generated events.

\begin{figure}[h!]
    \begin{center}
      \includegraphics[scale=0.45]{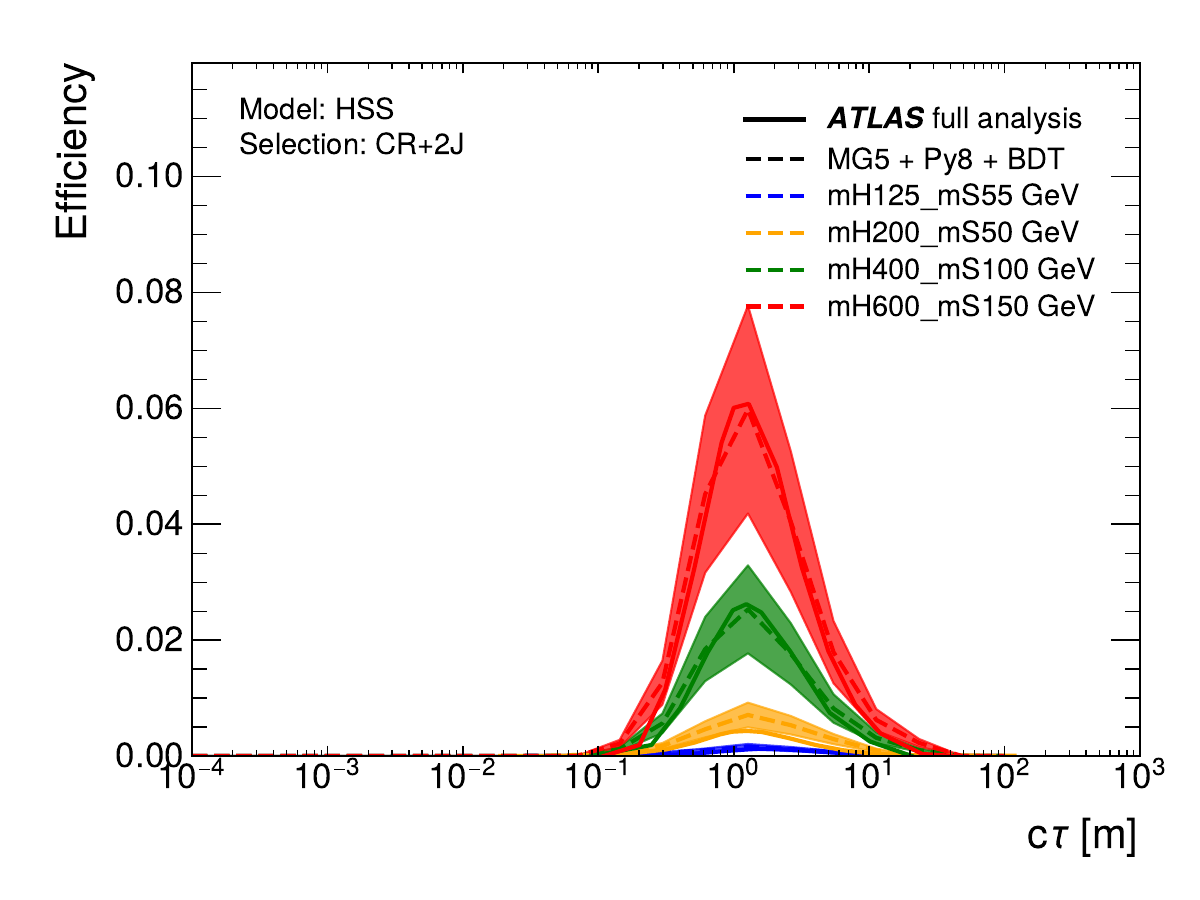} 
      
      \caption{\normalsize{Efficiencies obtained using the SuMos (dashed lines) compared to the results of the original ATLAS analysis (solid lines) for various signal parameter points, for the CR+$2J$ selection.} \label{comparison_CR+2J}}
  \end{center}
\end{figure}

The results for the CR+$2J$ selection in \autoref{comparison_CR+2J} show that the SuMo does an excellent job at reproducing the efficiency of the full analysis, over a range of Higgs-mediated signal samples, across the whole lifetime range. 

\begin{figure}[h!]
    \begin{center}
      \includegraphics[scale=0.45]{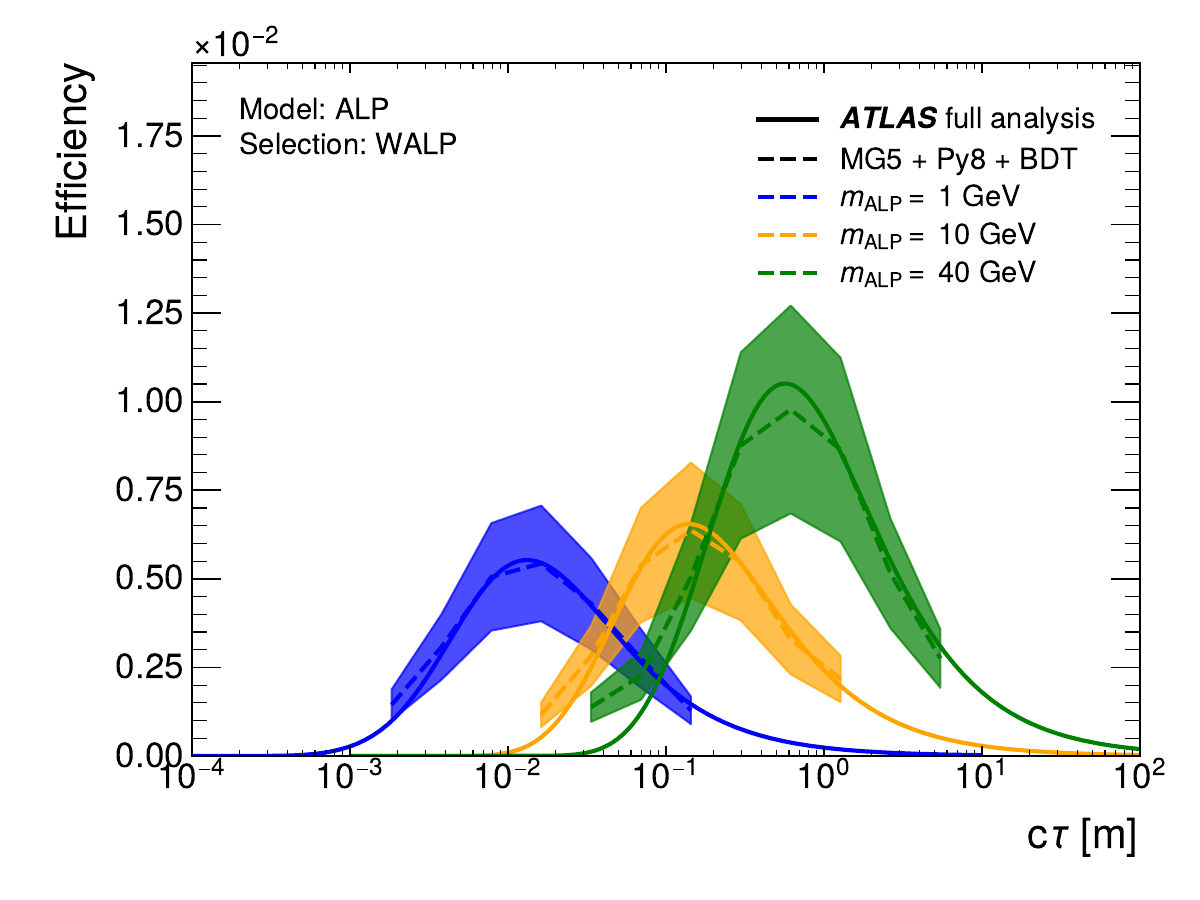} 
      
      \caption{\normalsize{Efficiencies obtained using the SuMos (dashed lines) compared to the results of the original ATLAS analysis (solid lines) for various signal parameter points, for WALP selection.} \label{comparison_WALP}}
  \end{center}
\end{figure}

\begin{figure}[h!]
    \begin{center}
      \includegraphics[scale=0.45]{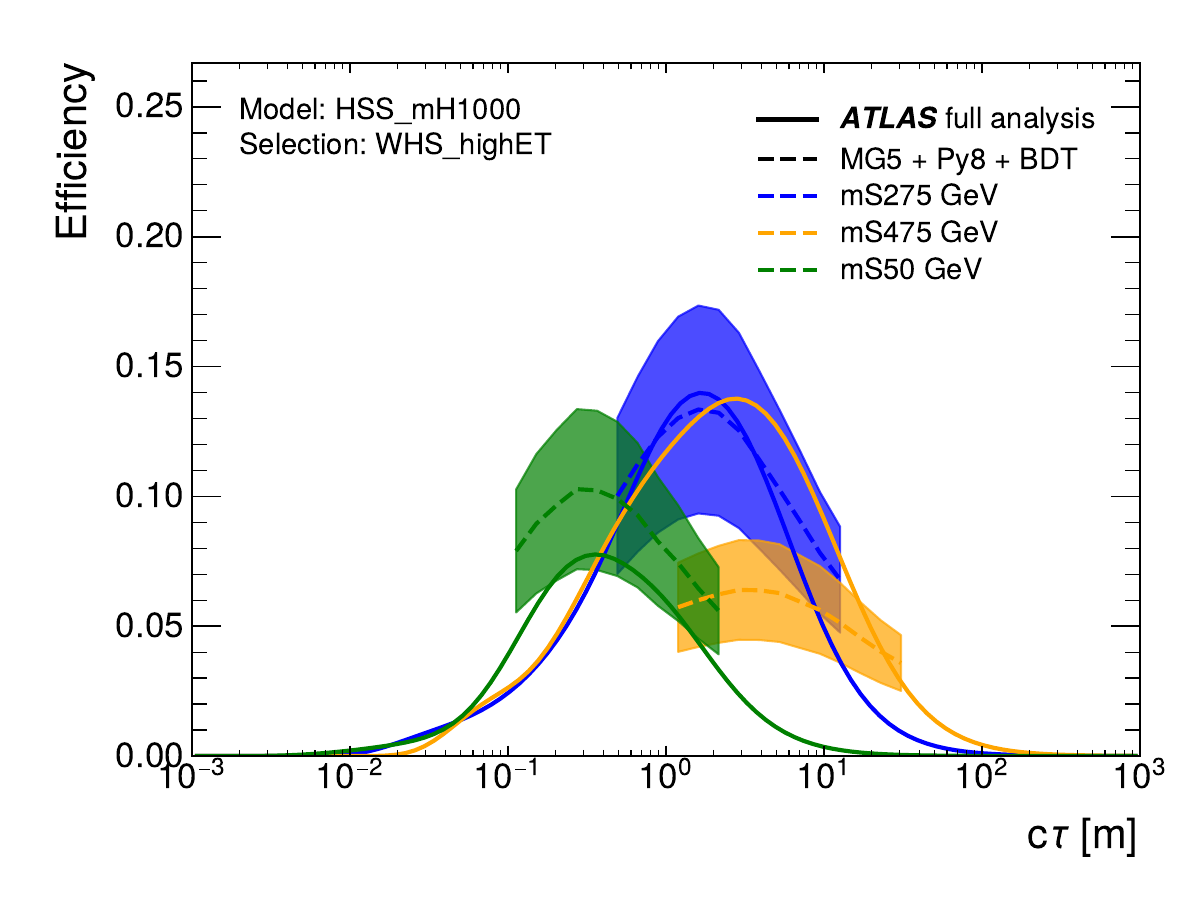} 
       \includegraphics[scale=0.45]{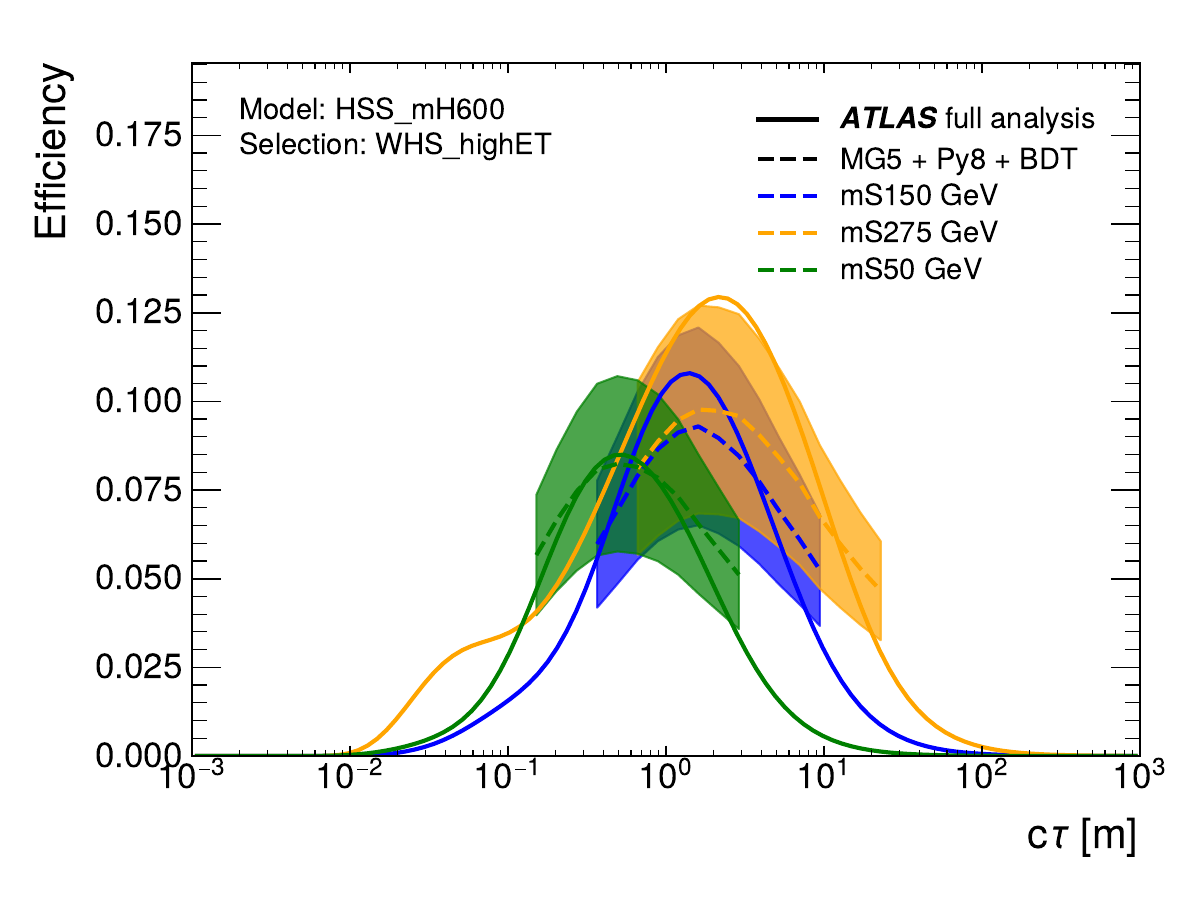} 
      \includegraphics[scale=0.45]{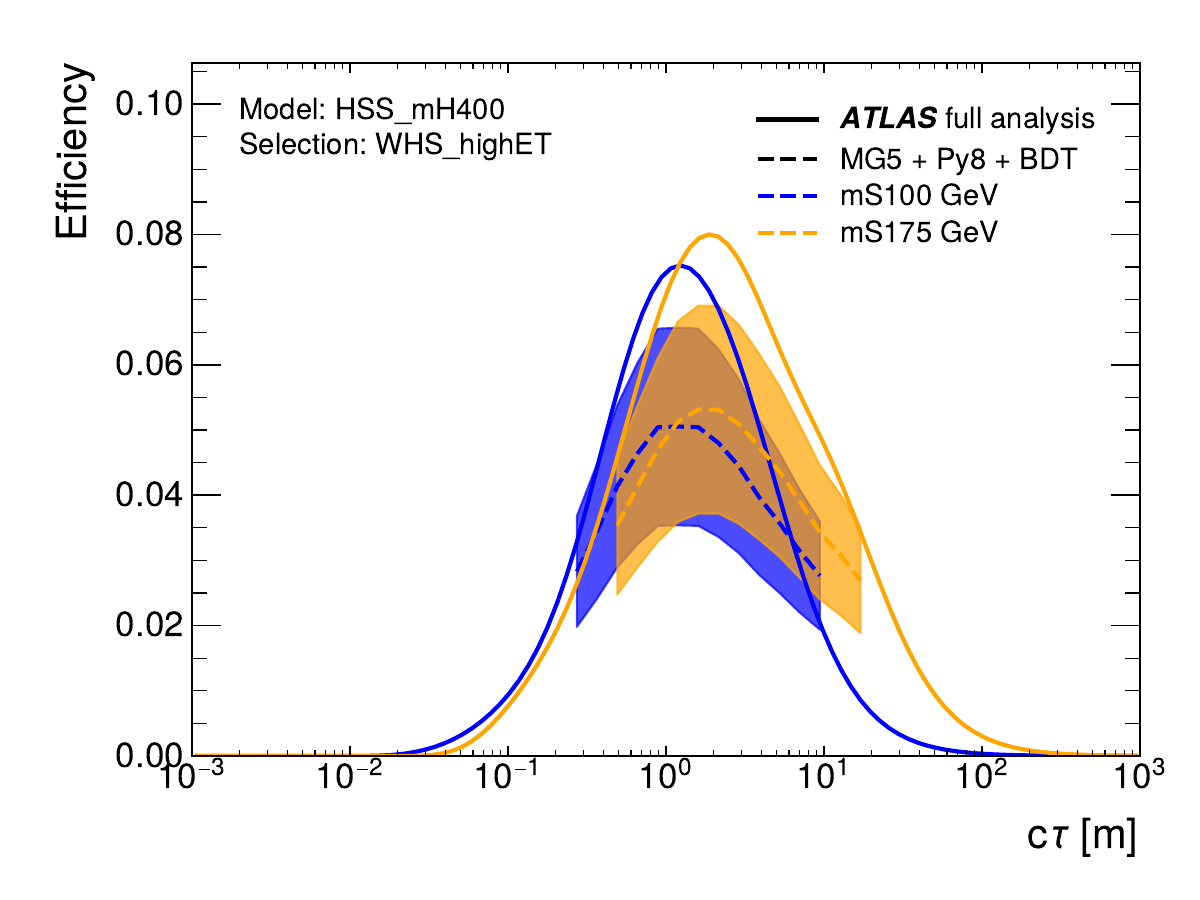} 

      \caption{\normalsize{Efficiencies obtained using the SuMos (dashed lines) compared to the results of the original ATLAS analysis (solid lines) for various signal parameter points, for the WHS\_highET selection.} \label{comparison_WHS_highET}}
  \end{center}
\end{figure}

The WALP selection results in \autoref{comparison_WALP} show efficiency curves for a variety of ALP mass scenarios. The SuMo in this case tends to slightly under-estimate the efficiency of the analysis but is consistent with the results from the original analysis within the 25\% uncertainty on the SuMo prediction, across the lifetime range where the SuMos are valid. The situation is a little more mixed for the WHS\_highET selection in \autoref{comparison_WHS_highET}, where the predictions in some cases over-shoot and others under-shoot, but typically within the the uncertainty band in the lifetime range where the model is deemed applicable.

\begin{figure}[h!]
    \begin{center}
          \includegraphics[scale=0.45]{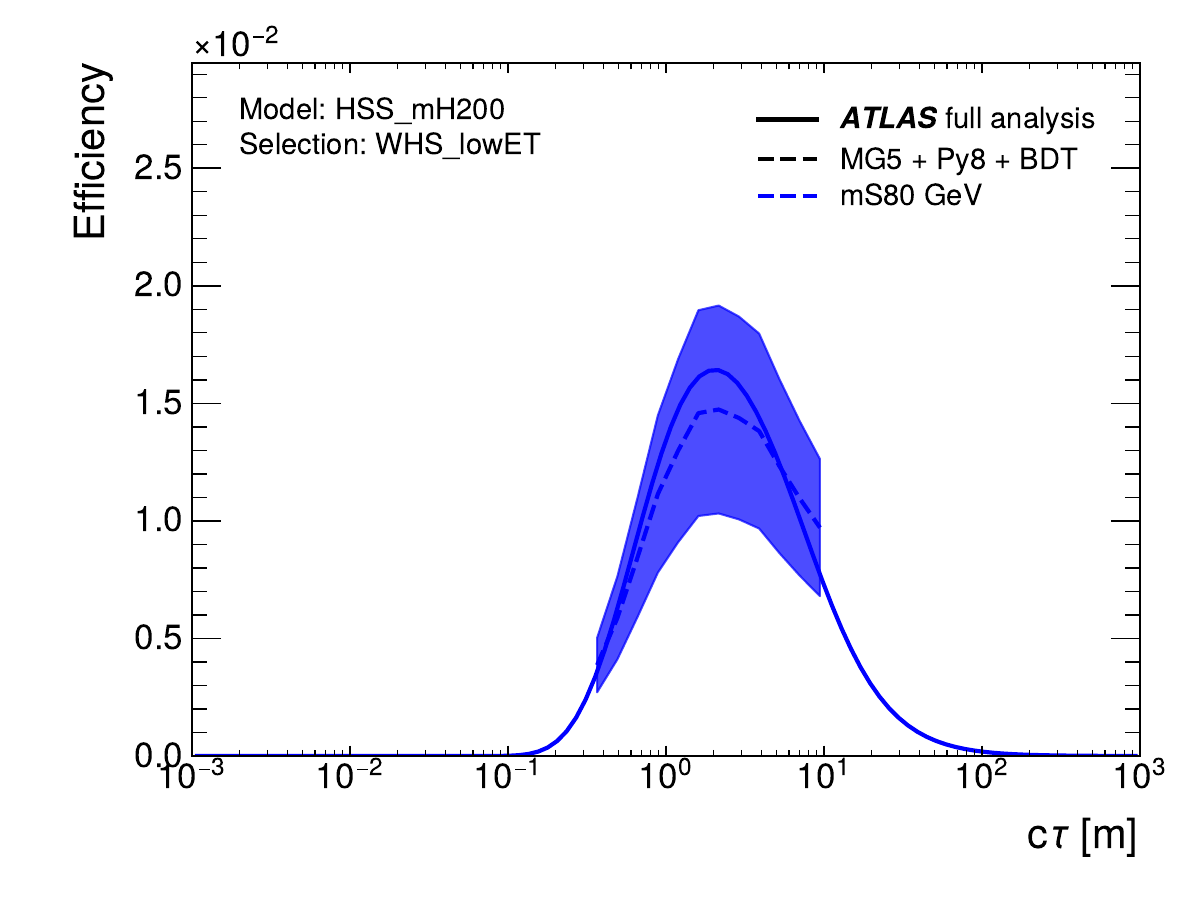} 
      \includegraphics[scale=0.45]{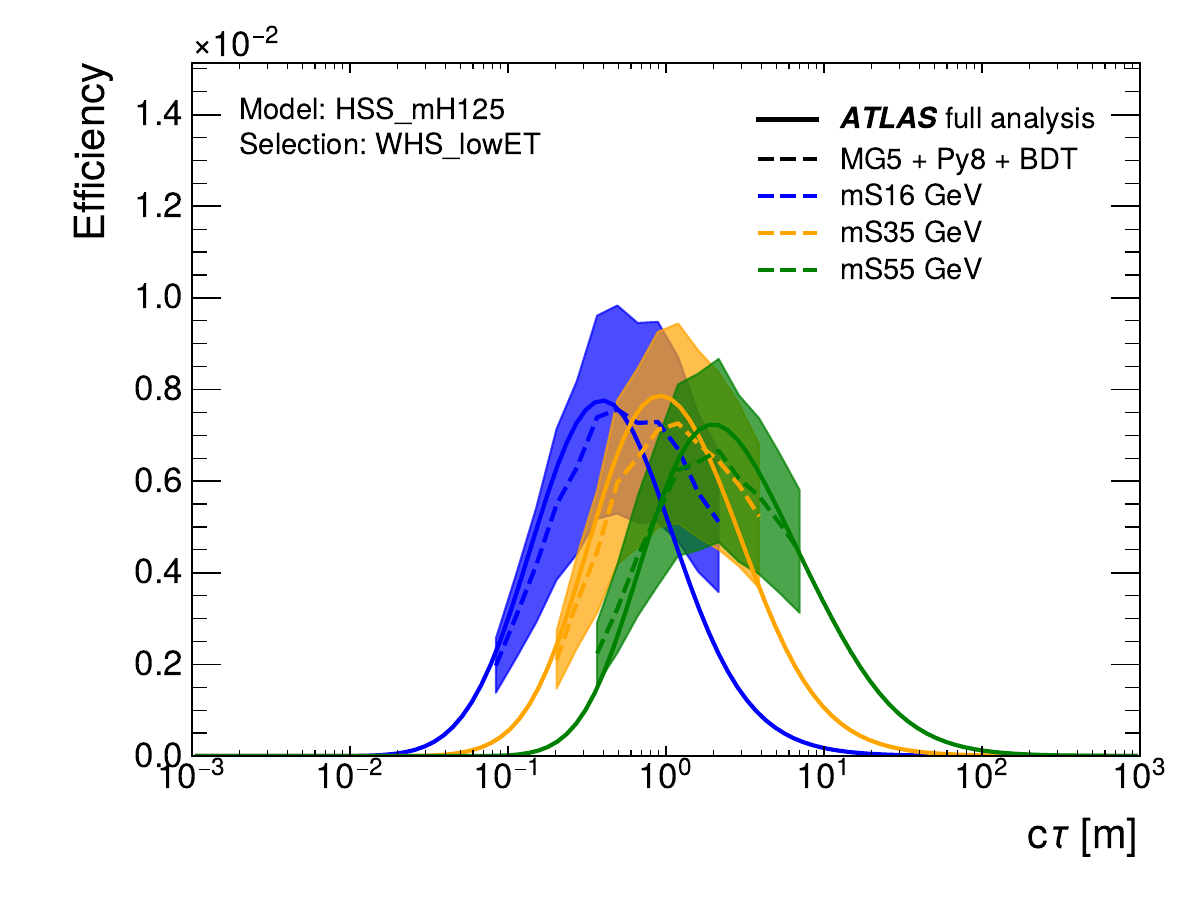} 

      \caption{\normalsize{Efficiencies obtained using the SuMos (dashed lines) compared to the results of the original ATLAS analysis (solid lines) for various signal parameter points, for the WHS\_lowET selection.} \label{comparison_WHS_lowET}}
  \end{center}
\end{figure}

One exception is the sample where the mediator mass is 1 TeV and the LLP mass is 475 GeV, and thus decays to top quarks. This efficiency appears to be largely under-estimated by the SuMo, by a factor of approximately 2. The results for the SuMo for WHS\_lowET shown in \autoref{comparison_WHS_lowET} show excellent agreement for all probed models. 

\begin{figure}[h!]
    \begin{center}
          \includegraphics[scale=0.45]{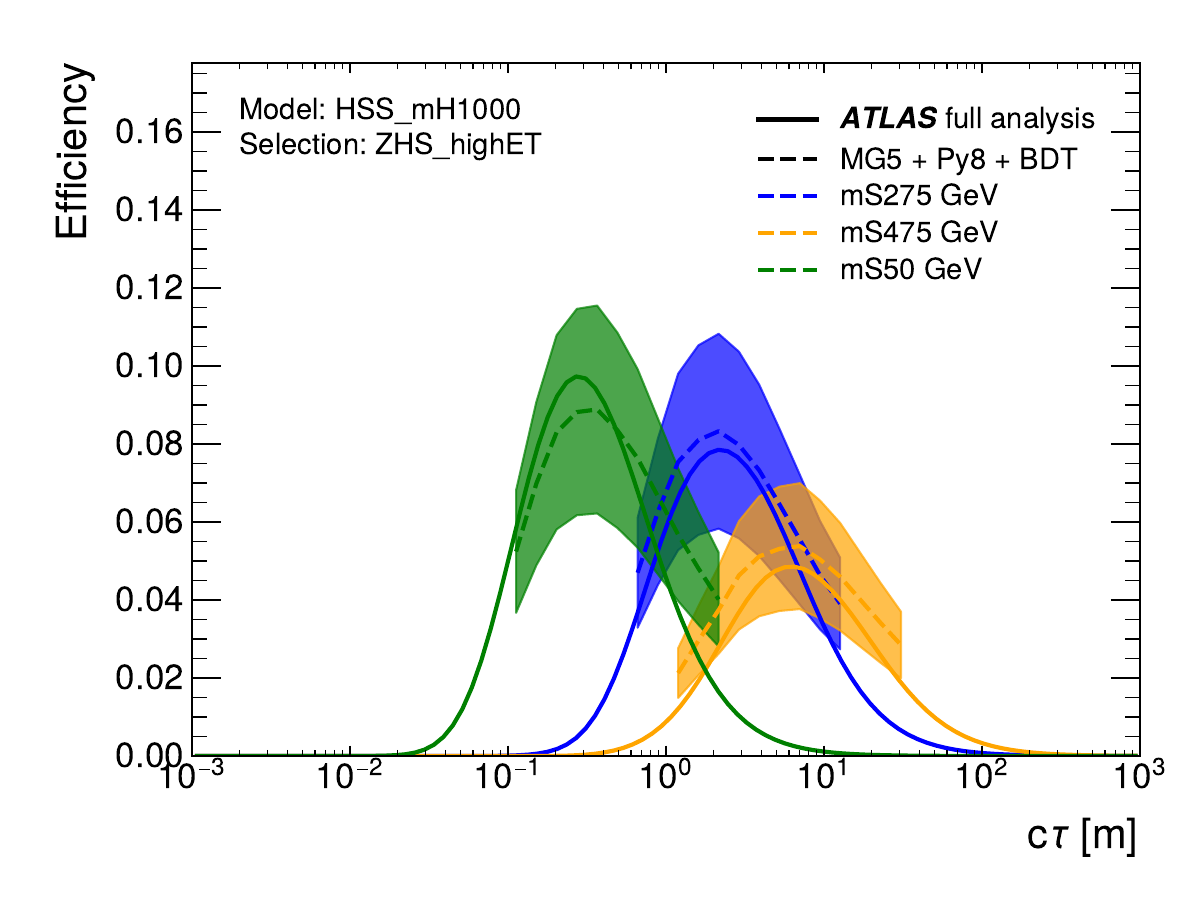} 
      \includegraphics[scale=0.45]{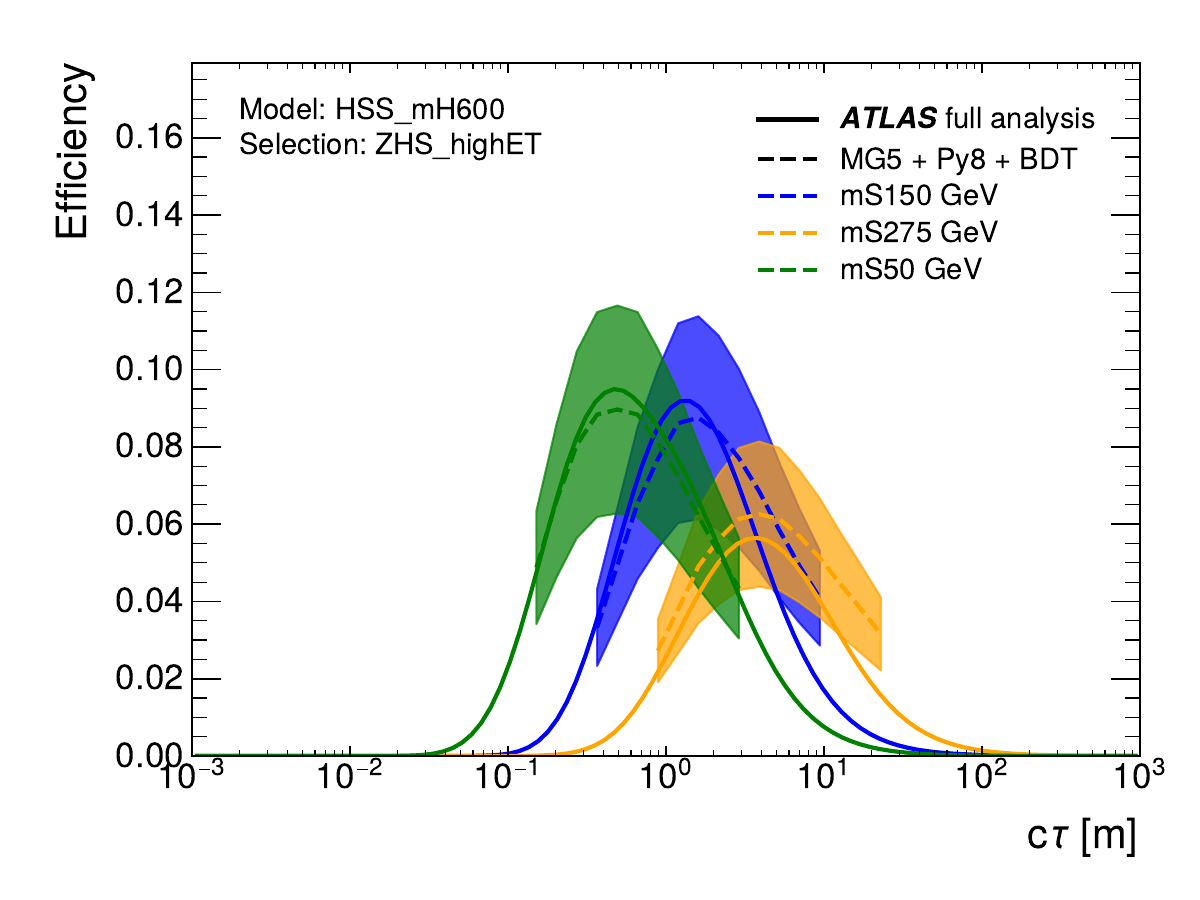} 
      \includegraphics[scale=0.45]{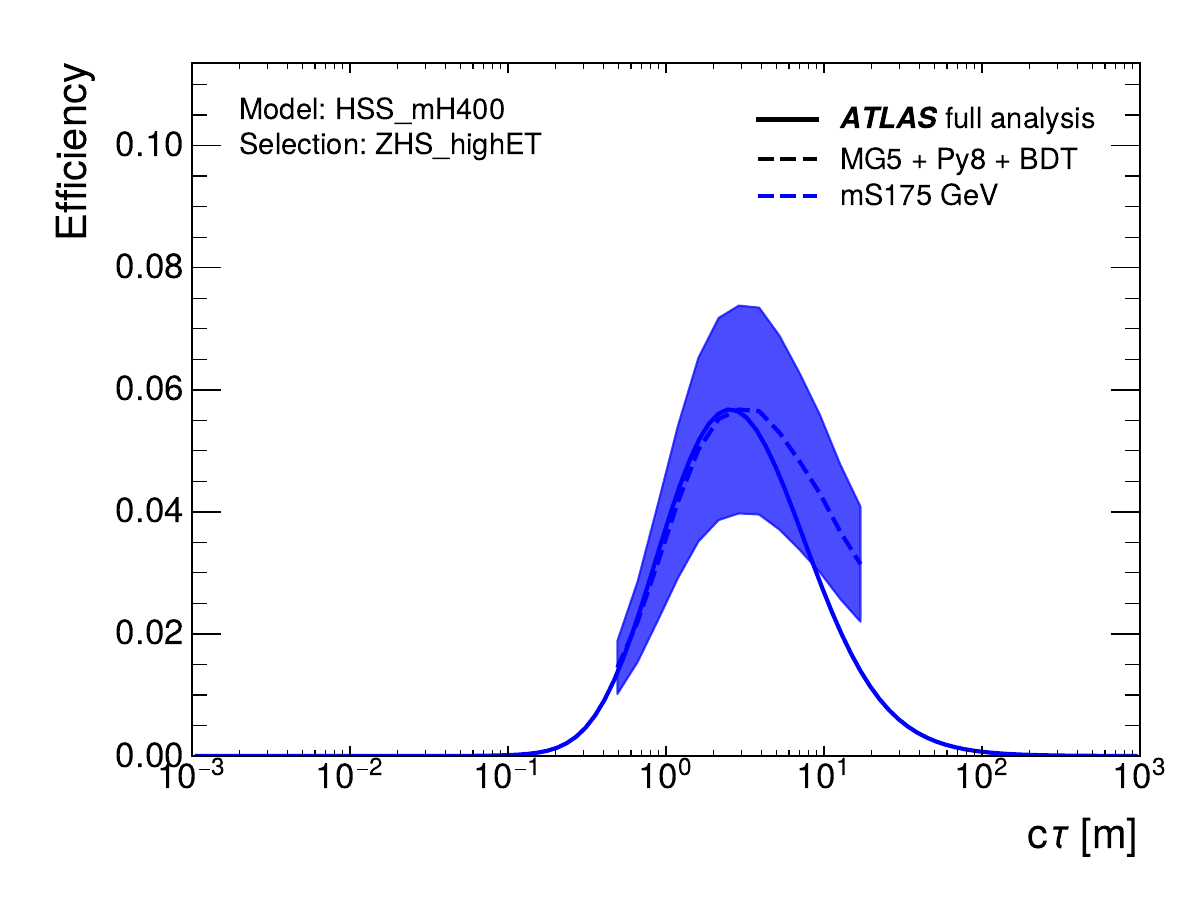}

      \caption{\normalsize{Efficiencies obtained using the SuMos (dashed lines) compared to the results of the original ATLAS analysis (solid lines) for various signal parameter points, for the ZHS\_highET selection.} \label{comparison_ZHS_highET}}
  \end{center}
\end{figure}

\begin{figure}[h!]
    \begin{center}
          \includegraphics[scale=0.45]{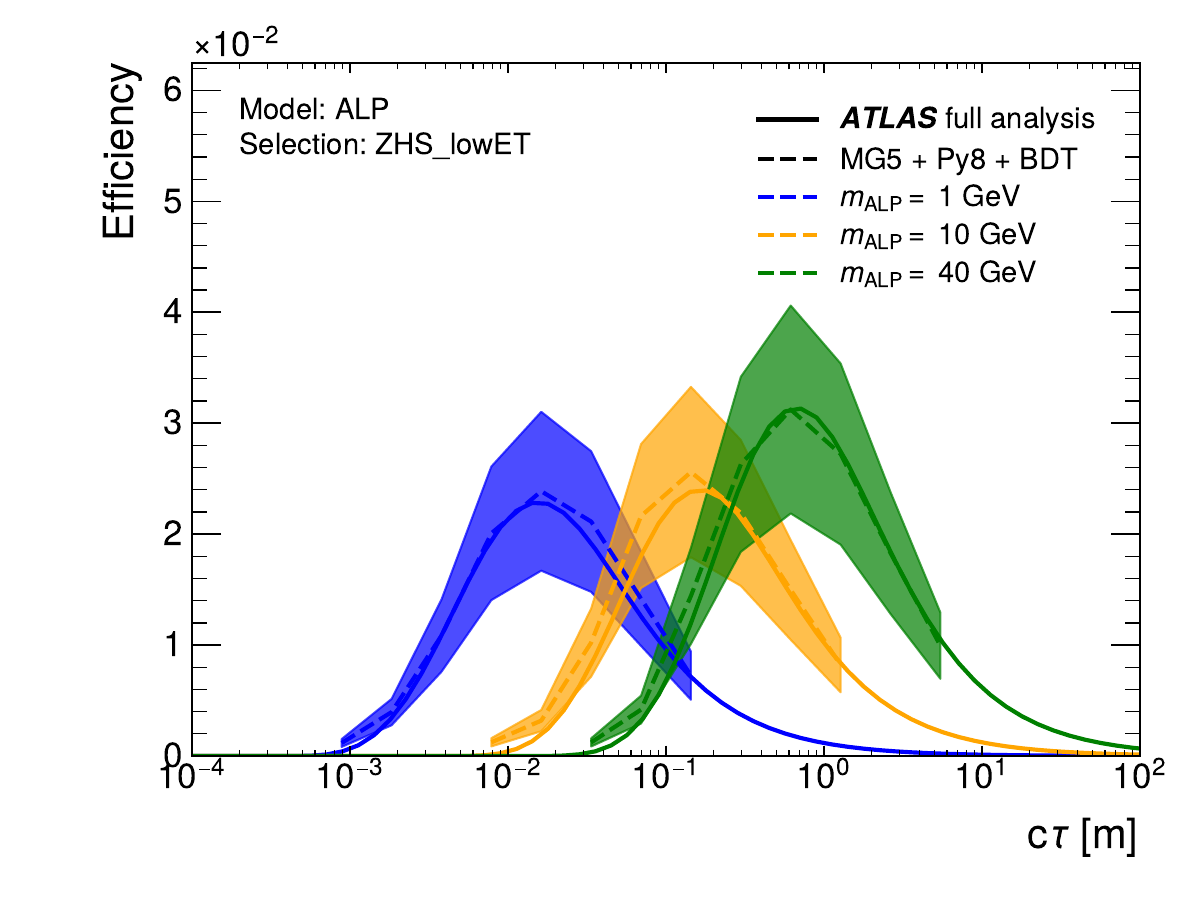} 
      \includegraphics[scale=0.45]{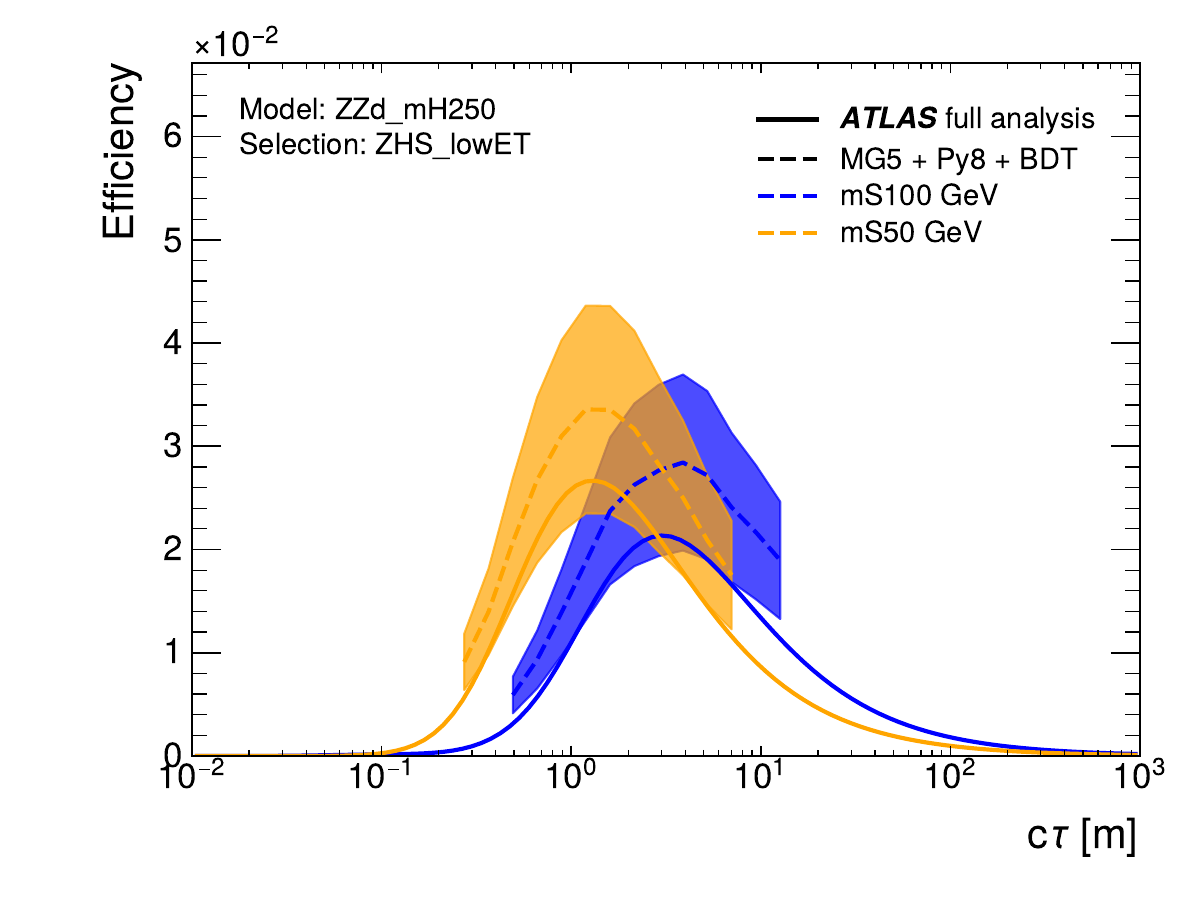} 
      \includegraphics[scale=0.45]{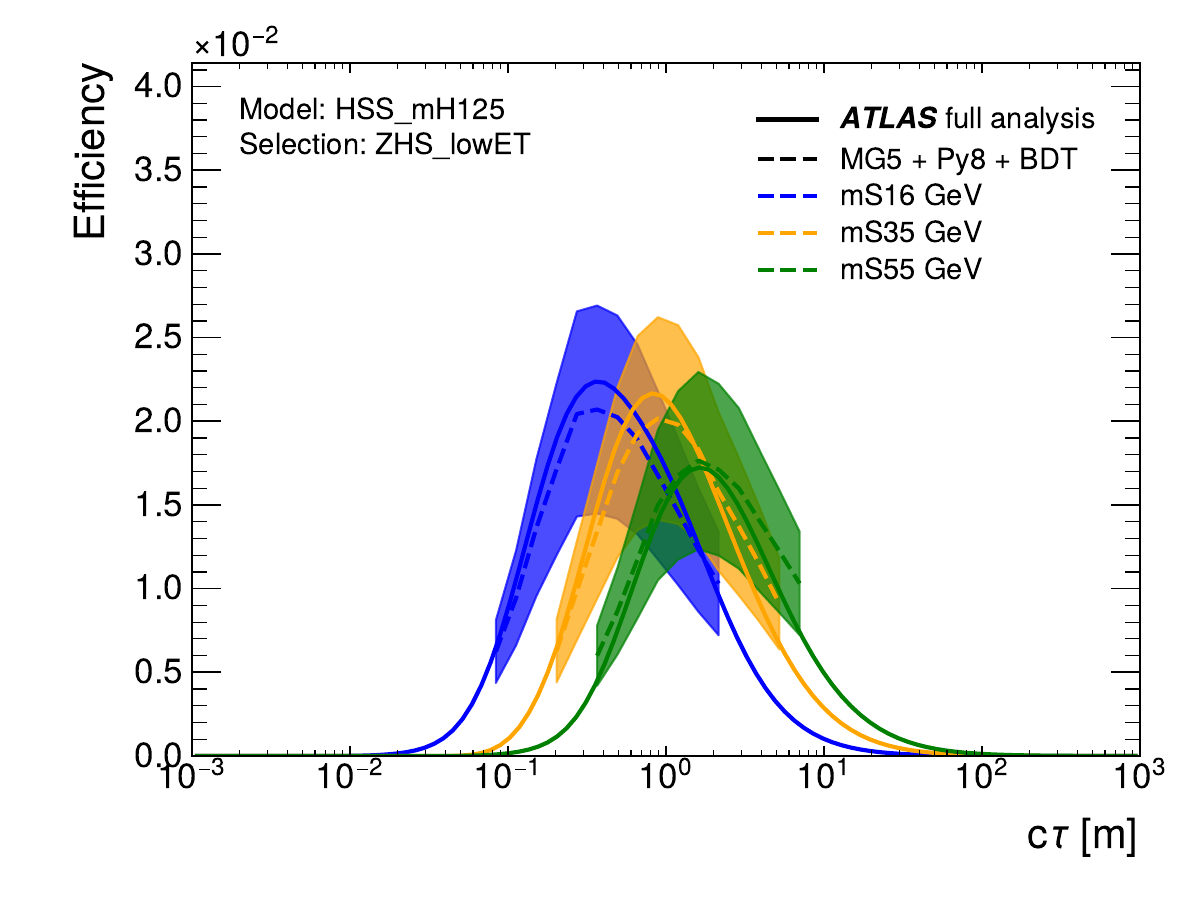} 
 
      \caption{\normalsize{Efficiencies obtained using the SuMos (dashed lines) compared to the results of the original ATLAS analysis (solid lines) for various signal parameter points, for the ZHS\_lowET selection.} \label{comparison_ZHS_lowET}}
  \end{center}
\end{figure}

The agreement from the predicted SuMo efficiency with the full analysis in the ZHS\_highET selection in \autoref{comparison_ZHS_highET} is excellent for all probed Hidden Sector models in the applicable lifetime range. Finally, the SuMo for ZHS\_lowET shown in \autoref{comparison_ZHS_lowET} show good agreement for all probed models within the 25\% uncertainty. This is quite impressive given that the models span Hidden Sector, Dark photon and ALP model types (which can decay to gluons or quarks depending on the model).

In conclusion, the performance of the SuMos is excellent over a wide range of models and final states, with a few exceptions where the prediction deviates from the 25\% uncertainty. In the most pathological cases, within the applicable $c\tau$ range, the SuMos still the correct efficiency within a factor of about two.

\section{\textbf{Validation using HackAnalysis}}
\label{hackana}

Since the SuMo method is substantially different to a conventional recast, implementation in standard recasting tools requires some accommodations. In a traditional recast, the particle level event information would be fed into a detector simulation, and this would typically involve keeping only the final state particles (this is true for both {\sc Delphes}~\cite{deFavereau:2013fsa} and {\sc SFS}~\cite{Araz:2021akd} detector approximation tools). Even in analyses with long-lived particles, typically only particles with a sufficient decay length are kept and made available to the recasting code, and usually the mother-daughter relationships are not retained (since that is not known to a real detector); for example, in the LLP implementations in {\sc MadAnalysis} described in Ref.\cite{Araz:2021akd}, LLP vertices were found by matching the decay positions of the LLP to the production vertices of their children. 
For the SuMo procedure, however, we require the truth-level information about the W/Z boson emitted, and the PDGID codes of the decay products of any LLPs. Jet clustering, detector simulation and such are superfluous. 

Therefore, to implement the SuMos in {\sc HackAnalysis}, a new ``detector'' routine was created (following the description in Ref.\cite{Goodsell:2024aig}) which retains all parent-child information about the event and keeps all fields, including very short-lived ones such as the $Z/W$ bosons, and this is made available to the analysis. No jet clustering for this routine is performed, so it is equivalent to reading the {\sc HEPMC} truth record, but it can accept either {\sc HEPMC} files or perform the decays/showering using {\sc Pythia8} internally from e.g. {\sc Les Houches Event} files. Some additions to the {\sc ONNX} interface were also required, because the BDT models available for this analysis contain more than one output node, and the relevant output node is a vector: it was necessary to allow the analysis to choose the desired node and element of the node to use for the efficiency. In the {\sc HackAnalysis} implementation each region is included as a cutflow, where the initial cuts first select those were there is at least one LLP, then classify events according to whether they have a $Z$ boson, $W$ boson, or two LLPs. The final cut is to apply the SuMo efficiency by reweighting the event region by region. 

The definition of an LLP for {\sc HackAnalysis} for this analysis had to be carefully chosen: in the procedure in the previous sections, the total efficiency for a \emph{sample} is taken to be zero if the \emph{average} decay of the particles is outside of certain ranges. In {\sc HackAnalysis} we must make this choice on an event-by-event basis, so the definition chosen is that an LLP must decay within the physical volume of ATLAS and within the tracker. In particular, it should not decay in a region less than $20$cm perpendicular to the beam pipe and $40$ cm along the beamline from the interaction point (these are roughly comparable to the region of validity of the SuMo). In initial tests, relaxing these constraints led to poor performance for low decay lengths and boosts. 

\begin{figure}\centering
\includegraphics[width=0.45\textwidth]{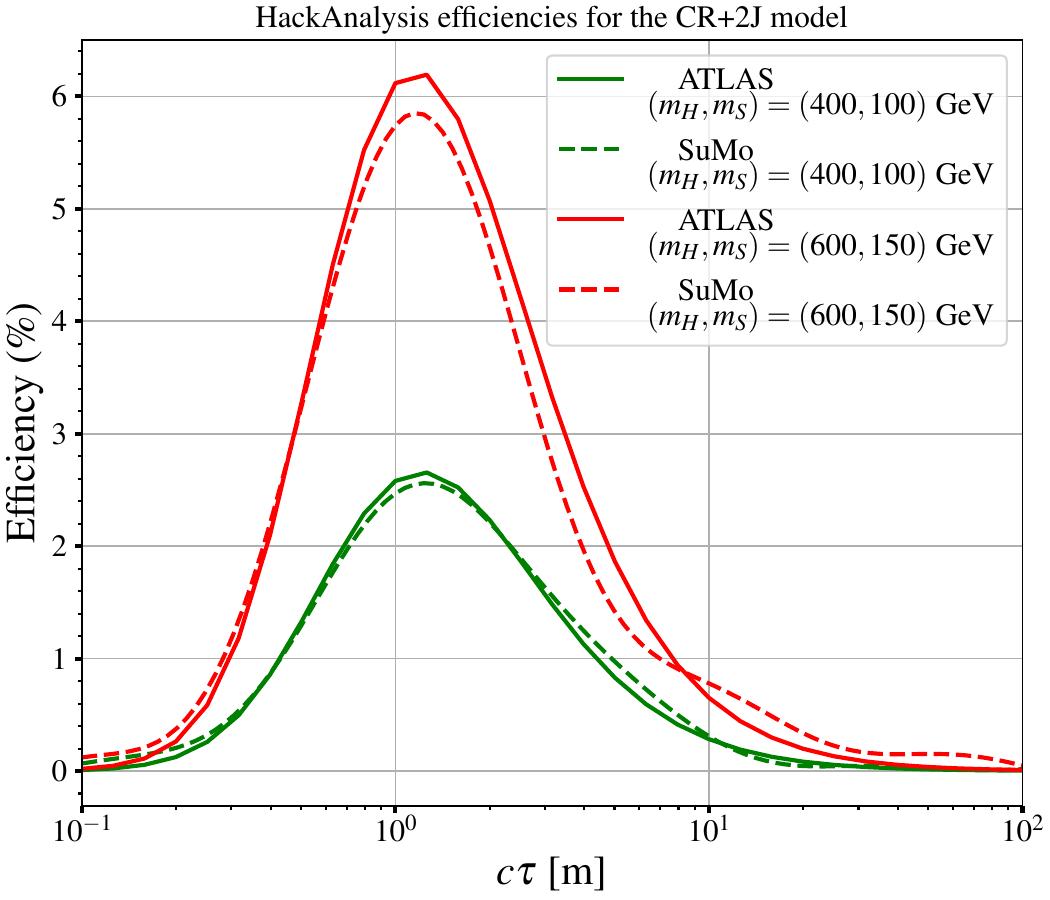}\caption{\label{FIG:HACR2J} Efficiencies obtained for the CR+2J region and ggH model using {\sc HackAnalysis}, compared to the full ATLAS results from the reference analysis~\cite{ATLAS-EXOT-2022-04}.}
\end{figure}

\begin{figure}\centering
\includegraphics[width=0.45\textwidth]{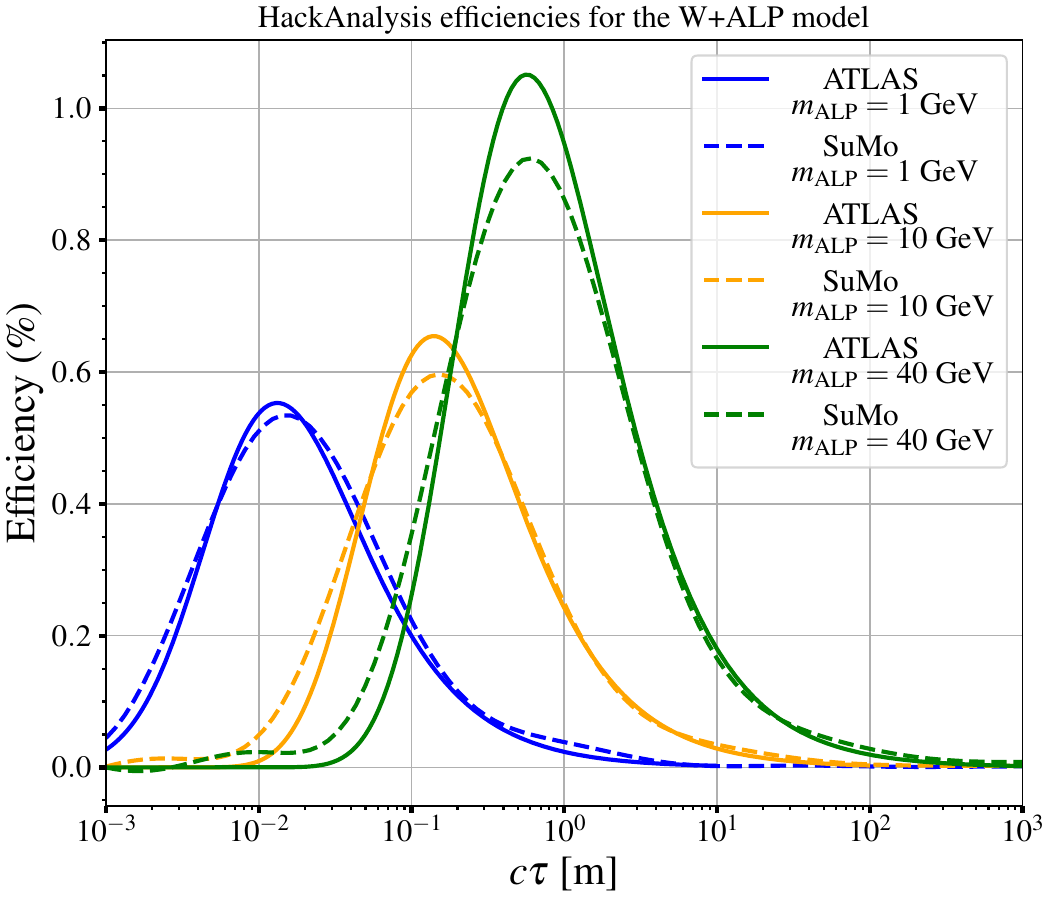}\caption{\label{FIG:HAWALP} Efficiencies obtained for the W+ALP model using {\sc HackAnalysis}, compared to the full ATLAS results from the reference analysis~\cite{ATLAS-EXOT-2022-04}.}
\end{figure}

This analysis is to be made available in {\sc HackAnalysis 2.3}. We validated the analysis using the W+ALP model on the WALP region, and the gluong-gluon fusion -produced HS model on the CR+2J region; the plots can be seen in Figures~\ref{FIG:HACR2J} and~\ref{FIG:HAWALP} and can be compared to Figures~\ref{comparison_CR+2J} and~\ref{comparison_WALP} respectively (running/scanning/plotting was handled using {\sc BSMArt} \cite{Goodsell:2023iac}). The agreement is excellent.

\section{\textbf{Pros and Cons of SuMos as analysis preservation tools}}

The studies performed in this validation note allow us to weigh up some pros and cons of the SuMo method as applied for the reference search~\cite{ATLAS-EXOT-2022-04}. 

We will start with the negatives. It is true that despite good overall performance across the model space checked here, there remain areas where disagreement is not within the stated error bands. Although this is only for some models, it does cause concern that one might not be able to fully trust the error bands when recasting some new model. This is partially relieved by the fact that the SuMos come with recommendations on which ranges of LLP lifetimes are appropriate.
And further, it is clear that although the SuMos do not give perfect agreement for all models, their results are typically within 25\% of the true value within the lifetime range where they claim to be valid. This is a considerable achievement given that the searches which they are emulating are practically impossible for theorists to reproduce with conventional recasting methods, on account of their heavy use of machine-learning algorithms using the detailed detector response as input. Hence, while they are not perfect, they open the door to non-trivial recasting of searches which would otherwise not be recastable \emph{at all}.

Another potential downside is that the SuMos were trained with relatively small number of trees (namely, 10). This implies that the probabilities returned for any given event are necessarily in increments of 0.1: this appears too coarse and could fail to correctly capture events with very low probabilities. A simple solution would be to train the models with a larger number of trees. Another easy-to-fix problem is that the zeroth output in the returned list is in fact the `not selected' category. This is counter-intuitive, and it would be better for the zeroth element to be the search region prediction.

Finally, one might be concerned by the fact that the models were trained on a relatively limited number of topologies, which might make it difficult to interpret the results when deviating from the original characteristics of the model. For example, all training models assumed decays of the LLPs to pairs of particles. If one wanted to test a model where the LLPs decayed to three quarks or gluons, it is not clear that the model would give an accurate estimate of the selection probability. A more problematic issue is that some decays, for example to light quarks instead of $b$-quarks, might obtain predicted efficiencies of zero since none of the reference models contained those decays... even if, on the face of it, one would expect some residual sensitivity. Hence, when using these SuMos one must not blindly trust the output, but instead make informed judgements about their applicability and possible generalisations. This judgement should come from scrutinizing the details of the original search and judging the overlap of a new model with the assumptions of the final states of the benchmark models. This is the downside of the fact that the BDTs are a ''black box'' unlike a conventional cutflow-based recasting where each step of the selection is mastered.

Now, we focus on some of the biggest upsides of the models. The first one to note is that because they run on truth-level inputs, the costly and time-consuming task of detector simulation can be avoided, since it is folded into the SuMo. 
This advantage was already true for the ``efficiency map'' objects which were validated in Ref.~\cite{ThomasNote}, but the SuMos present several advantages. In particular, by generalising the maps into machine-learning structures, an immediate benefit is that the SuMos can be trivially encapsulated in ONNX objects. This format is designed to be portable, and is already well-used in the phenomenological community, facilitating the easy adoption of the SuMos. This replaces the clunky yaml-based format in which the efficiency maps were preserved. For example, the easy portability of the SuMos as ONNX objects allowed them to be almost immediately ported into the \textsc{HackAnalysis} framework~\cite{Goodsell:2024aig}. A related advantage in terms of usage is that the ONNX model is easy and fast to evaluate, compared to efficiency maps where the identification of the relevant value to retrieve for each event was a slow process, and error-prone. When used in python, the whole operation can also be vectorized, bringing further performance gains.

The use of BDTs also has other operational advantages compared to custom objects like efficiency maps, for the analysis teams preparing them. Indeed, it means that additional variables can be trivially added: for instance in the SuMos under discussion, the fact that the transverse mass and the transverse momentum are both included in the feature list means that the SuMo can infer any dependence of the result on the mass difference between the parent and daughter particles, helping to improve performance. This would not have been trivial for effiemcy map objects. Similarly, the efficiency maps discussed in Ref.~\cite{ThomasNote} suffered from sub-optimal binning choices for the input features, which could significantly affect the performance, since regions where important dependencies could exist may have been lumped into a single bin. The BDT structure of these SuMos obviates that problem, since the ``bins'' are replaced by a series of decision trees, whose cuts are adapted to give the best performance. Finally, while providing predictions for multiple regions (aside from the signal region A) would have been structurally unfeasible with an efficiency map object, this comes ``for free'' with a BDT using the ``multi-class'' feature: effectively the algorithm is trained to report 5 values instead of 1, such that predictions for all regions can be obtained in one shot and with no additional difficulties for the user. This becomes important in models where a significant number of signal events could end up in other regions, thus necessitating their inclusion in the statistical analysis. 

In summary, the SuMos appear to open the door to preservation of analyses which were otherwise impossible to recast, while adhering the to FAIR principles of open science:
\begin{itemize}
    \item Findeable -- since can be found via the usual HEPData portal page for the reference analysis;
    \item Accessible -- because they are freely downloadable and provided in a well-known format;
    \item Inter-operable -- because the ONNX format is already designed to work in various programming languages (notably C++ and python), across multiple platforms (Unix, Windows,...) and are already integrated in major recasting frameworks (HackAnalysis, Rivet,...)\footnote{An implementation in {\sc MadAnalysis}\cite{Conte:2012fm,Conte:2014zja,Conte:2018vmg} should be straightforward once the ONNX interface is publicly available.};
    \item Re-useable -- because they allow analysis results to be re-used for purposes beyond their original benchmark models.
\end{itemize}


\section{\textbf{Summary}}

This note provided an independent validation of the surrogate model objects provided as re-interpretation material by a recent ATLAS search for displaced jets. We find that the surrogate models provide a very good approximation of the full analysis efficiency using only truth-level information. Furthermore, these models have several operational advantages both for the analysts and downstream users. This opens the door to rolling out these types of objects as a method of preserving experimental analyses for other LLP searches, and indeed for other searches.

\addtolength{\textheight}{-0cm}   




\printbibliography

\end{document}